# Single-particle Mass Spectrometry with arrays of frequency-addressed nanomechanical resonators

Eric Sage[1,2], Marc Sansa[1,2], Shawn Fostner[1,2], Martial Defoort[1,2], Marc Gély[1,2], Akshay K. Naik[3], Robert Morel[4,5], Laurent Duraffourg[1,2], Michael L. Roukes[6], Thomas Alava[1,2], Guillaume Jourdan[1,2], Eric Colinet[1,2,†], Christophe Masselon[7,8,9], Ariel Brenac[4,5], Sébastien Hentz[1,2*]

1. Univ. Grenoble Alpes, F-38000 Grenoble, France
2. CEA, LETI, Minatec Campus, F-38054 Grenoble, France
3. Centre for Nano Science and Engineering, Indian Institute of Science, Bangalore, 560012, India
4. Univ. Grenoble Alpes, INAC-SP2M, F-38000 Grenoble, France
5. CEA, INAC- SP2M, F-38000 Grenoble, France
6. Kavli Nanoscience Institute and Departments of Physics, Applied Physics, and Bioengineering, California Institute of Technology, MC 149-33, Pasadena, California 91125 USA
7. CEA, IRTSV, Biologie à Grande Echelle, F-38054 Grenoble, France
8. INSERM, U1038, F-38054 Grenoble, France
9. Université Joseph Fourier, Grenoble 1, F-38000, France

ABSTRACT

One of the main challenges to overcome to perform nanomechanical Mass Spectrometry (NEMS-MS) analysis in a practical time frame stems from the size mismatch between the analyte beam and the extremely small nanomechanical detector area. We report here the demonstration of NEMS-MS with arrays of 20 individually addressed nanomechanical resonators where the number of inputs/ouputs for the whole array is the same as that of a single resonator. While all resonators within an array are interconnected via two metal levels, each resonator is designed with a distinct resonance frequency which becomes its individual "address". In order to perform single-particle Mass Spectrometry, the resonance frequencies of the two first modes of each NEMS within an array are monitored simultaneously. Using such an array, mass spectra of metallic aggregates in the MDa range are acquired with more than one order of magnitude improvement in analysis time due to the increase in capture cross section compared to individual resonators. A 20 NEMS array is probed in 150ms with the same mass limit of detection as a single resonator. Spectra acquired with a conventional Time-Of-Flight (TOF) mass spectrometer in the same system show excellent agreement. As individual information for each resonator within the array is retained, the array becomes a particle imager, each resonator acting as a pixel. With this technique, we demonstrate how Mass Spectrometry Imaging (MSI) at the single particle level becomes possible by mapping a 4cm-particle beam in the MDa range and above.

TEXT

Mass Spectrometry (MS) has been one of the fastest-growing analytical techniques over the past two decades[1,2] and has become an essential tool in a broad variety of fields[3–5]. MS is particularly well suited to the analysis of light molecules: it is based on ionization, which raises issues for high-mass species[6]. Routine use of MS in the MDa (~1.66 ag) to GDa (~1.66 fg) range remains challenging: while currently out of reach for commercial instruments, a few specialized systems have shown the ability to study supramolecular assemblies in the one to tens of MDa mass range[7,8]. In this mass range, interesting results have been recently obtained with unconventional MS architectures like Charge Detection systems[9,10] also based on ionization of species. In parallel, mass sensing using nanomechanical resonators has been performed for the last fifteen years with a variety of devices[11] and a mass limit of detection of a few Daltons has been reported using a carbon nanotube[12]. Nano-Electro-Mechanical Systems (NEMS) operate best in the MDa to GDa range, and real-time acquisition of NEMS-MS for single proteins was demonstrated with top-down silicon resonators[13]. An important milestone has been reached recently with the demonstration of MS of particles regardless of their charge with NEMS[6], which can circumvent issues associated with ionization of species, in particular transfer efficiency. The excellent mass limit of detection obtained with nanomechanical devices comes at the cost of an extremely reduced capture cross-section, and this is of course all the more true for bottom-up devices like carbon nanotubes. One of the main challenges to overcome to perform NEMS-MS analysis in a practical time frame stems from the size mismatch between the analyte beam and the nanomechanical detector area[13,14]. It is therefore of crucial importance to significantly increase the capture cross-section of resonators while maintaining their outstanding performance. Capturing a larger proportion of particles will also decrease the amount of sample required to perform an analysis, which can be a requirement for some biological species. In the past, gas sensing was demonstrated with dense and large arrays of identical interconnected NEMS[15]. In this case, gas molecules adsorb homogeneously onto the surface of all NEMS within the array, which operate collectively and simultaneously. This is not suitable for NEMS-MS, as information about each single device is lost in the collective operation of the array: a single particle would shift the frequency of only one device, and this information would be averaged over the whole array.

We report here the demonstration of NEMS-MS with arrays of individually addressed nanomechanical resonators where the number of inputs/ouputs for the whole array is the same as that of a single resonator. While all resonators within an array are interconnected via two metal levels, each resonator is designed with a distinct resonance frequency which becomes its individual "address". NEMS within an array operate in multi-mode[6] and retain the same mass resolution as a single resonator. Using such an array, mass spectra of metallic aggregates have been acquired with excellent speed due to a significantly enhanced capture cross section compared to individual resonators. Spectra acquired with a conventional Time-Of-Flight (TOF) mass spectrometer in the same system showed excellent agreement. As individual information for each resonator within the array was retained, we could demonstrate spatial imaging of a particle beam at the single particle level.

We used monocrystalline silicon resonators fabricated from Silicon-On-Insulator wafers with Very Large Scale Integration processes[16]. The resonators were electrostatically actuated and used a differential piezoresistive readout. Two asymmetric drive electrodes were used to simultaneously operate each resonator on two resonance modes and deduce both mass and position of accreted particles in real time[6]. Every input/output of all resonators within an array was interconnected across resonators: for instance, the output pad of the array was electrically connected to the output pads of all resonators within the array. With five inputs/outputs per resonator, this was only made possible by using two metal levels and vertical interconnects (see Figure 1a to d). The number of electrical pads for the whole array was thus the same as that for a single resonator (see Figure 1f). One obvious advantage of such a configuration is that the same measurement setup could be used for both single resonators and arrays (provided a sufficiently high measurement bandwidth), without the need for complex wire-bonding or addressing. Fabrication details can be found in Supplementary Information.

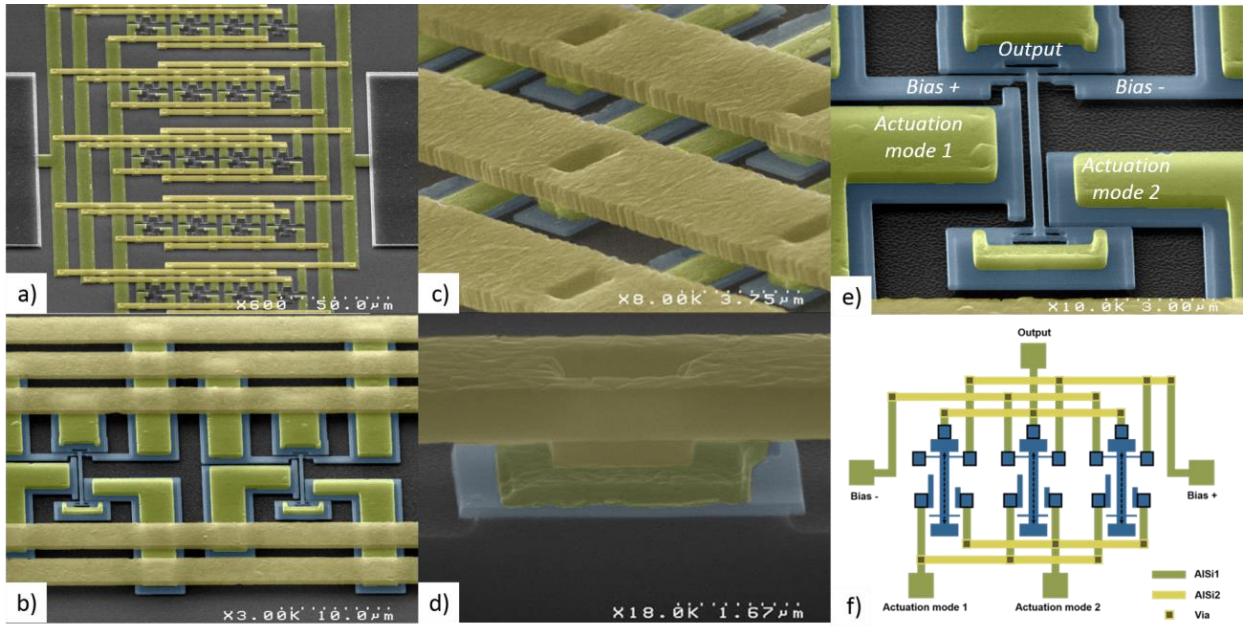

*Figure 1:* ***Array of nanomechanical resonators***. *SEM images of 5×4 NEMS array used for nanomechanical mass spectrometry. Typical horizontal and vertical pitches are 20 and 55μm respectively. a) General view of the array b) zoom on two resonators (silicon is false-coloured in deep blue), and their metal interconnects (AlSi). c) and d) zoom-in on interconnects and via. The first metal level is coloured in deep yellow, the second one in light yellow. e) Typical doubly clamped in-plane resonator used in this study. The beams were designed to resonate around 30 MHz for mode 1 and 80 MHz for mode 2. Typical dimensions for the resonant beam were: 160 nm (thickness), 300 nm (width) and 5–10 μm (length). In-plane motion transduction was performed using piezoresistive nanogauges in a bridge configuration to allow background cancellation. Electrodes were specifically patterned for efficient mode 1 and mode 2 actuation. For a resonance frequency $f_0$, bias voltages at $f_0 + \Delta f$ were applied to both nanogauges (with 180° dephasing). Tension/compression in the gauges mix their resistance change to obtain a downmixed differential output voltage at $\Delta f$, typically around a few 10's of kHz. f) Schematic of the interconnect layout. Each resonator has a unique beam length, hence a unique resonance frequency.*

Retrieving individual information corresponding to each resonator within the array as required by NEMS-MS was performed by "frequency addressing": Distinct resonance frequencies for each resonator were obtained by slightly varying their length. The frequency pitch needed to be large enough so that resonance peaks did not overlap after fabrication due to process uncertainty as well as after mass deposition which results in downshifts in resonance frequency. The choice of pitch was a trade-off between risk of spectrum overlap, number of resonators within the array and measurement bandwidth. Arrays of 20 (5*4) resonators were fabricated with typical spatial pitches of 20 μm and 55 μm in the horizontal and vertical direction respectively. Resonance frequencies of our arrays typically ranged from 20 to 45 MHz for mode 1, and 70 to 120 MHz for mode 2 with frequency pitches in the 500 kHz range (length difference in the 150 nm range).

An open loop frequency sweep response of the array in vacuum (~$10^{-5}$ Torr) and at liquid nitrogen temperature was performed with a downmixing scheme (see Figure 2a and measurement details in Supplementary Information). Resonance peaks were found to be well separated from each other with excellent Signal-to-Background ratios (~55dB), around 180° phase shifts and quality factors ($Q$) very similar to typical $Q$s of individual resonators[6] (*e.g.* 8500 and 7500 in average for the first and second mode respectively). This is despite the fact that for a given drive voltage, the output signal of the array at a given beam's resonance frequency scales like $\frac{1}{N-1}$, $N$ being the number of resonators within the array (see Supplementary Information). While the frequency pitch was designed to be constant, a spread was observed due to fabrication uncertainties.

Our arrays were operated in closed loop to monitor the resonance frequencies of all resonators sequentially over time (see Figure 2b): initial frequencies and phase references were recorded for every NEMS. A Phase Lock Loop (PLL) subsequently locked onto a given resonator and registered a frequency data point after a given idling time $\tau_{PLL}$, before switching to the next resonator. The duty cycle of a whole array was therefore $N\tau_{PLL}$ where $N$ is the number of probed NEMS. Each data point was subsequently assigned to its corresponding resonator and individual time frequency traces could be deduced. Mass determination of a single particle requires the use of the second resonance mode[6]. The same procedure could be simultaneously performed with a second measurement channel (see Supplementary Information) and an additional PLL. Both first and second mode frequencies could be thus simultaneously monitored in real time and our arrays of nanomechanical resonators can be used to perform NEMS-MS.

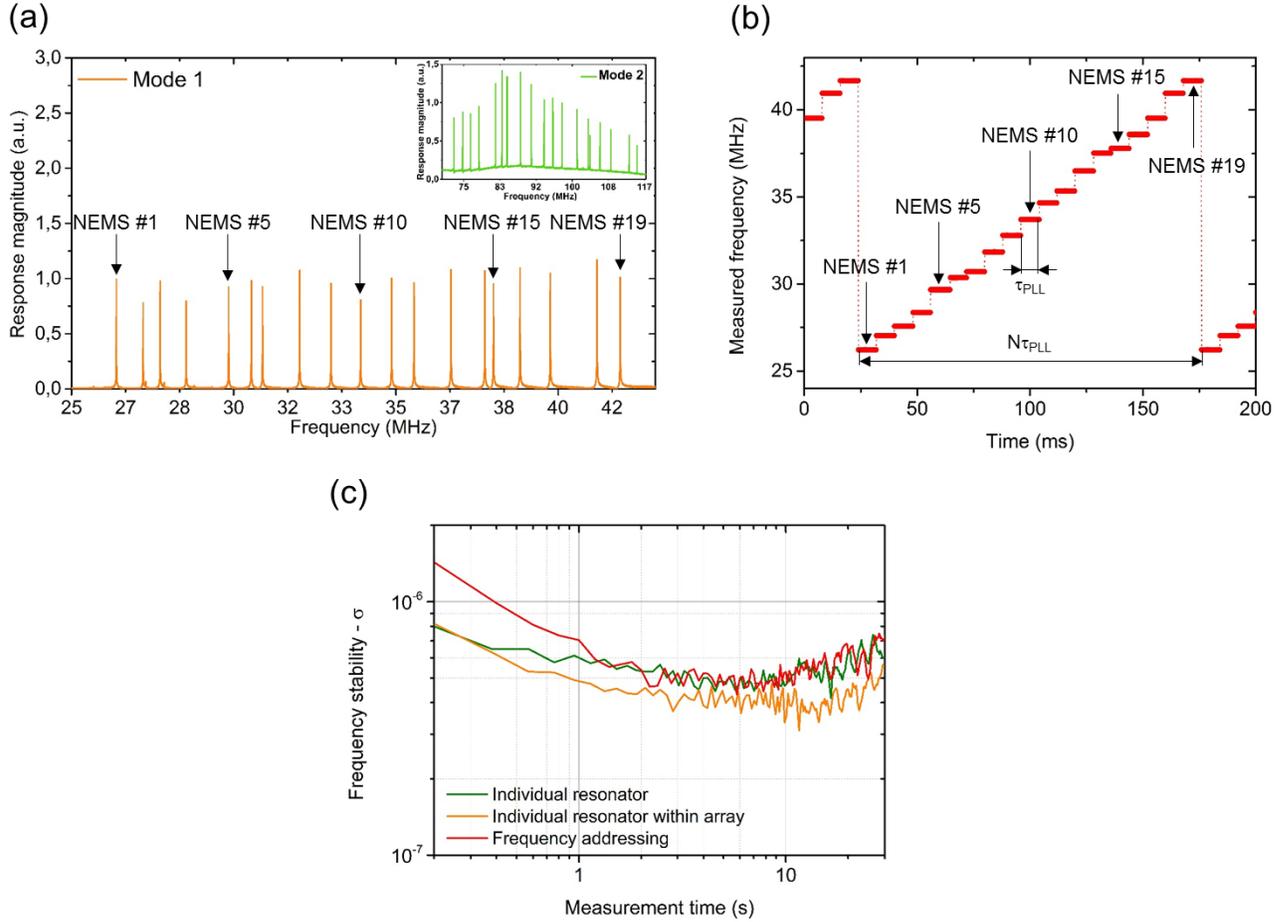

*Figure 2: **Frequency-addressing technique**. a) An open loop response of an array of 20 NEMS is recorded for mode 1 and mode 2 (inset). Each peak corresponds to the resonance of a single NEMS resonator for which resonance frequency and phase reference can be used as an address. We are showing here an example with only 19 resonance peaks: one resonator in the array failed after a long period of operation, as confirmed by Scanning Electron Microscopy (SEM) observation. Yet, the array as a whole could still be operated without performance degradation, demonstrating the robustness of the parallel architecture. b) The resonance frequency of every single resonator in the array is sequentially monitored over time: a PLL locks onto a given resonator, registers its current resonance frequency after a given idling time $\tau_{PLL}$ (here 8 ms) and then switches to the next resonator. The duty cycle of a whole array is then $N\tau_{PLL}$ (here 152 ms with N=19 NEMS). From the recorded data points, individual frequency time traces can be extracted, and their frequency stability calculated. c) Frequency stabilities obtained using a single individual resonator (not in array, green), a resonator of strictly identical dimensions within an array without frequency addressing (yellow) and the same resonator with frequency addressing (red). See Supplementary Information for details on the selected frequency stability estimator. The three plots appeared identical within measurement uncertainty: the parallel architecture of our arrays along with the frequency addressing technique allowed reaching the regime where frequency fluctuations set the frequency stability limit of our resonators, down to similar values as single resonators[17].*

Frequency stability is a key parameter to the performance of nanoresonators for mass sensing. In a regime where additive white noise is dominant, based on the simple dynamic range equation, the frequency stability can be expressed in the voltage domain as[17]:

$$\langle \frac{\delta f}{f_0} \rangle \cong \frac{1}{2Q} \frac{S_n}{V_{out}} \sqrt{BW} \qquad (1)$$

where $Q$ is the resonator's quality factor, $V_{out}$ the output signal amplitude at any given NEMS resonance frequency (in $V$), $S_n$ the noise spectral density at the output (in $V\ Hz^{-1/2}$), and BW the measurement bandwidth (in $Hz$). As mentioned above, the output signal of an individual resonator in array configuration scales like the inverse of the number of devices, $V_{out} \propto \frac{1}{(N-1)}$. In our case, $S_n$ is the sum of lock-in input noise (constant, typically less than $10 nV\ Hz^{-1/2}$), of Johnson noise due to the piezoresistive gauges and of thermomechanical noise. All piezoresistances being connected in parallel, Johnson noise scales like $\frac{1}{\sqrt{N}}$. Since Johnson noise for a single device is typically in the same order of magnitude as the lock-in input noise, it becomes negligible for an array. Thermomechanical noise for a single resonator is typically in the same order as Johnson noise or the lock-in input noise. Like output signal, it scales in the voltage domain like $\frac{1}{(N-1)}$ and becomes negligible for an array. Finally $S_n$ is dominated by the lock-in input noise, which does not scale with the number of resonators in the array, while the output voltage $V_{out}$ is inversely proportional to the latter. In a regime where additive white noise is dominant, the frequency stability of resonators within our arrays is expected to degrade proportionally with the number of resonators ($\langle \frac{\delta f}{f_0} \rangle \propto (N-1)$).

From the individual time frequency traces measured with our sequential closed-loop scheme, the frequency stability of every resonator within the array can be plotted. Figure 2c compares three different cases: the first is the frequency stability of a single resonator (not in an array) with a usual down-mixing scheme. The second is the frequency stability of a resonator of identical dimensions within an array, but operated with the same readout scheme (no frequency addressing). Finally, the third trace corresponds to the same resonator within an array with frequency addressing. The three plots are similar within measurement uncertainty. Yet, in the case of additive white noise, we could expect a factor 20 between the two first cases. We attribute this discrepancy to the presence of resonance frequency fluctuations in the mechanical domain: we have recently shown that the frequency stability of similar silicon single resonators was limited by frequency fluctuations rather than additive white noise[17]. In this regime, the frequency stability does not depend on signal level and depends very weakly on integration time. We found the same behavior with our arrays and using the frequency addressing technique. It should also be noted that frequency stabilities of all resonators within the array were of very similar levels (see Supplementary Figure S4). Finally, the frequency addressing technique did not degrade the frequency stability of our nanoresonators and the limits of detection in mass of single and arrayed resonators were similar.

We subsequently performed single-particle mass spectrometry with our arrays of nanomechanical resonators in a custom setup described in detail elsewhere[6,18]. The setup consisted of four main vacuum chambers (see Figure 3a): a metallic nanocluster source, an intermediate chamber, a deposition chamber and an in-line TOF mass spectrometer. Metallic nanoclusters were generated using a sputtering gas aggregation technique with tunable size and deposition rate. Nanoclusters were then expelled into the vacuum deposition chamber ($10^{-5}$ Torr) through a differential pumping stage. The deposition rate was measured using a quartz crystal microbalance (QCM) placed on a translational stage. Upon retraction of this stage, the array of resonators was exposed to the cluster flux. When both NEMS and QCM were retracted, the particle flux could enter the acceleration region of the in-line TOF mass spectrometer, where the mass-to-charge distribution of charged particles was measured. The configuration of the deposition chamber allows QCM, TOF-MS and NEMS-MS measurements sequentially on the same cluster population.

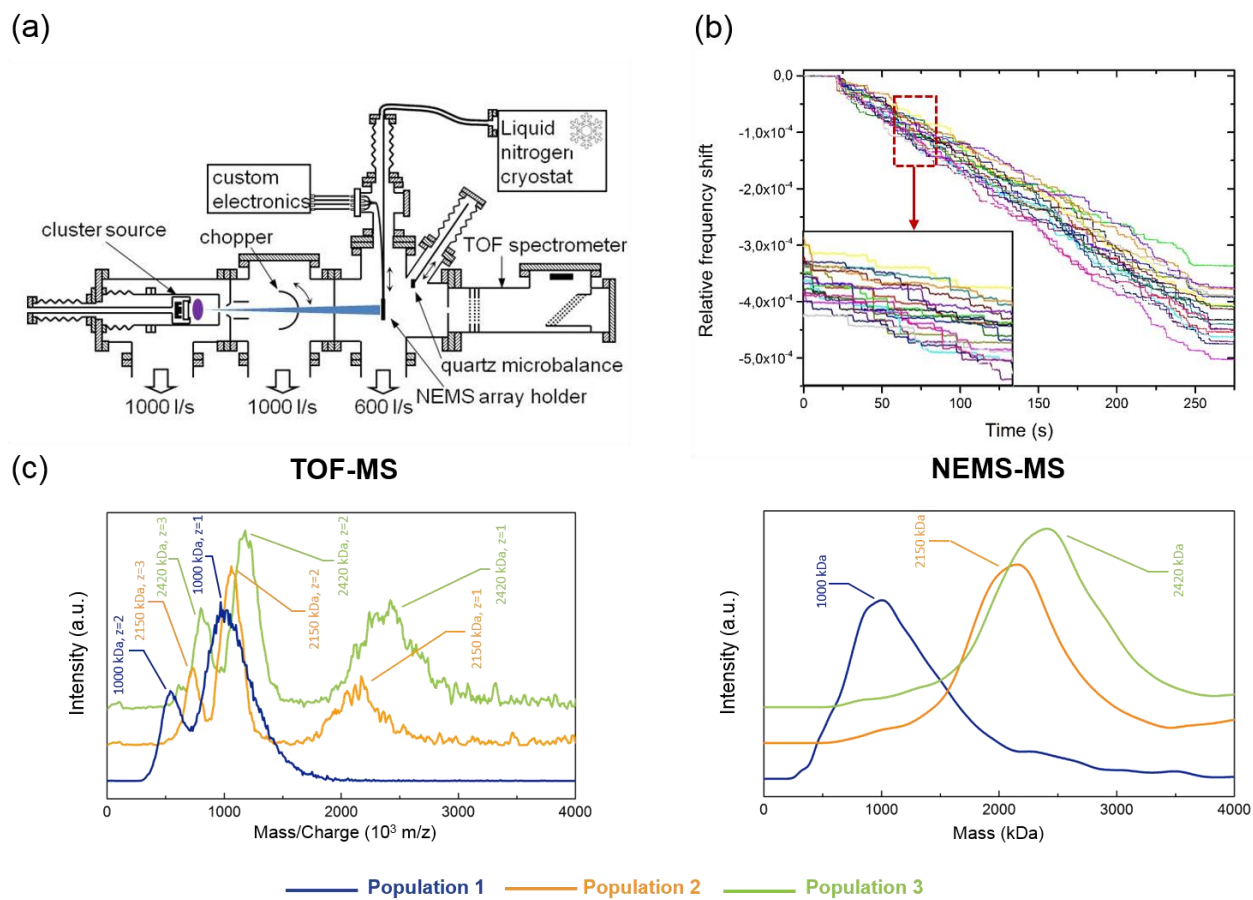

*Figure 3: **Single-particle Mass Spectrometry with arrays of nanoresonators** a) Schematic of the setup showing from left to right: the cluster source, an intermediate chamber containing a chopper, the deposition chamber and an in-line TOF mass spectrometer. Both NEMS holder and QCM were retractable, allowing for sequential NEMS-MS, TOF-MS and QCM measurements with the same operating conditions. b) Mode 1 relative frequency time traces of an array of 19 NEMS exposed to a flux of tantalum nanoclusters with a mean diameter of 7.2 nm. Inset: zoom-in with frequency jumps induced by single particle deposition c) Comparison of TOF and NEMS-MS with an array of 19 resonators performed with three distinct populations of nanoclusters with mean diameters of 5.8 nm (~1000 kDa), 7.4 nm (~2150 kDa) and 7.7 nm (~2420 kDa) respectively.*

As previously described[6], we selected tantalum as the analyte as it is both dense (16.6 g.cm$^{-3}$) and readily condenses into large clusters. The TOF and NEMS-MS mass spectra acquired for various populations were compared. An example of frequency traces in simultaneous two-mode operation acquired using the frequency addressing technique during the exposition to cluster beam is shown Figure 3b. Frequency jumps larger than the frequency stability 3σ were considered as actual particle landing events and converted into a mass probability distribution[6] (the stiffness of the particles was neglected due to their very small size, the in-plane motion of the resonator and its width-to-thickness ratio[14]) . The mass sensitivity of each NEMS was measured by comparing its frequency response to uniform mass deposition with that of a Quartz Crystal Microbalance (QCM) as detailed elsewhere[6]. This was performed for the 19 resonators simultaneously with the frequency addressing technique (see Supplementary Information). Extracted mass sensitivities ranged from 15.3 to 32.5 Hz/ag for mode 1 and 42.9 to 87.8 Hz/ag for mode 2, which was consistent with the range of the resonator lengths. Monitoring the first two modes of all NEMS was achieved with a PLL response time of 8 ms, yielding a total array response time of 152 ms. Tuning of the nanocluster source parameters and use of a mechanical chopper (see Figure 3a) yielded particle adsorption event rates per resonator in the order of one event every few seconds, making the landing of several particles within the duty cycle very unlikely. The mass probability distributions obtained for each event were added for each resonator to build individual mass spectra; the same operation could be performed for all resonators to build the overall array mass spectrum. Mass spectra of three different nanocluster populations acquired by both NEMS-MS and TOF-MS technique are displayed in Figure 3c. Just like individual nanomechanical resonators[6], NEMS-MS performed with arrays directly provided the cluster mass distribution independently of the particles charge state. Conversely, TOF-MS provided mass-to-charge ratio distributions corresponding to multiple charge states of the measured clusters, making spectra interpretation less straightforward. Each NEMS-MS spectrum was acquired in only 4 minutes and yielded approximately 1000 events. Each resonator detected a similar number of events during this amount of time (~50 events per resonator), demonstrating the 19-fold improvement in capture efficiency due to the use of the array. The overall spectrum provided an accurate mean mass of the cluster populations over a large mass range (530 kDa to 2400 kDa), with a broader distribution than the TOF spectrometer. As a matter of fact, these experiments were performed in a mass range compatible with operation of the TOF mass spectrometer, *i.e.* just above the resonator's mass limit of detection. Over a few MDa, ions are not sufficiently accelerated in order for the ion detector to provide a signal and the TOF spectrometer becomes unable to perform a correct analysis (see Supplementary Figure S6). Conversely, the NEMS limit of detection remaining constant with mass, its resolving power (ratio of analyzed mass to mass resolution) improves with increasing mass (see Supplementary Figure S7). For a given cluster population however, arrays yielded slightly broader peaks than those of a single resonator. We attribute this to the heterogeneity in both mass sensitivities and mass resolutions of individual NEMS across the array (see Supplementary Figures S8 and S9). This effect will become negligible at masses far from this limit of detection, where mass resolution will become negligible compared to measured mass. Nonetheless, our results demonstrated that such frequency-addressed arrays multiply the capture efficiency by the number of individual resonators in the array, in our case, by more than an order of magnitude.

The frequency addressing scheme also provided access to individual information of each resonator. This could be put to use, for example, to obtain a spatial mapping of the particle beam.

For this purpose, the 100 μm* 250 μm NEMS array could be moved to scan the 4 cm-diameter particle beam. Figure 4 shows maps of event number within the array in a given measurement time (here, 4 min), as well as individual spectra obtained with each resonator. These results are presented for two different array locations within the particle beam: i) close to its center, where the event rate is very homogeneous throughout the array and ii) at the edge of the beam, where there is a clear asymmetry between resonators situated well within the particle beam and the resonators outside of it.

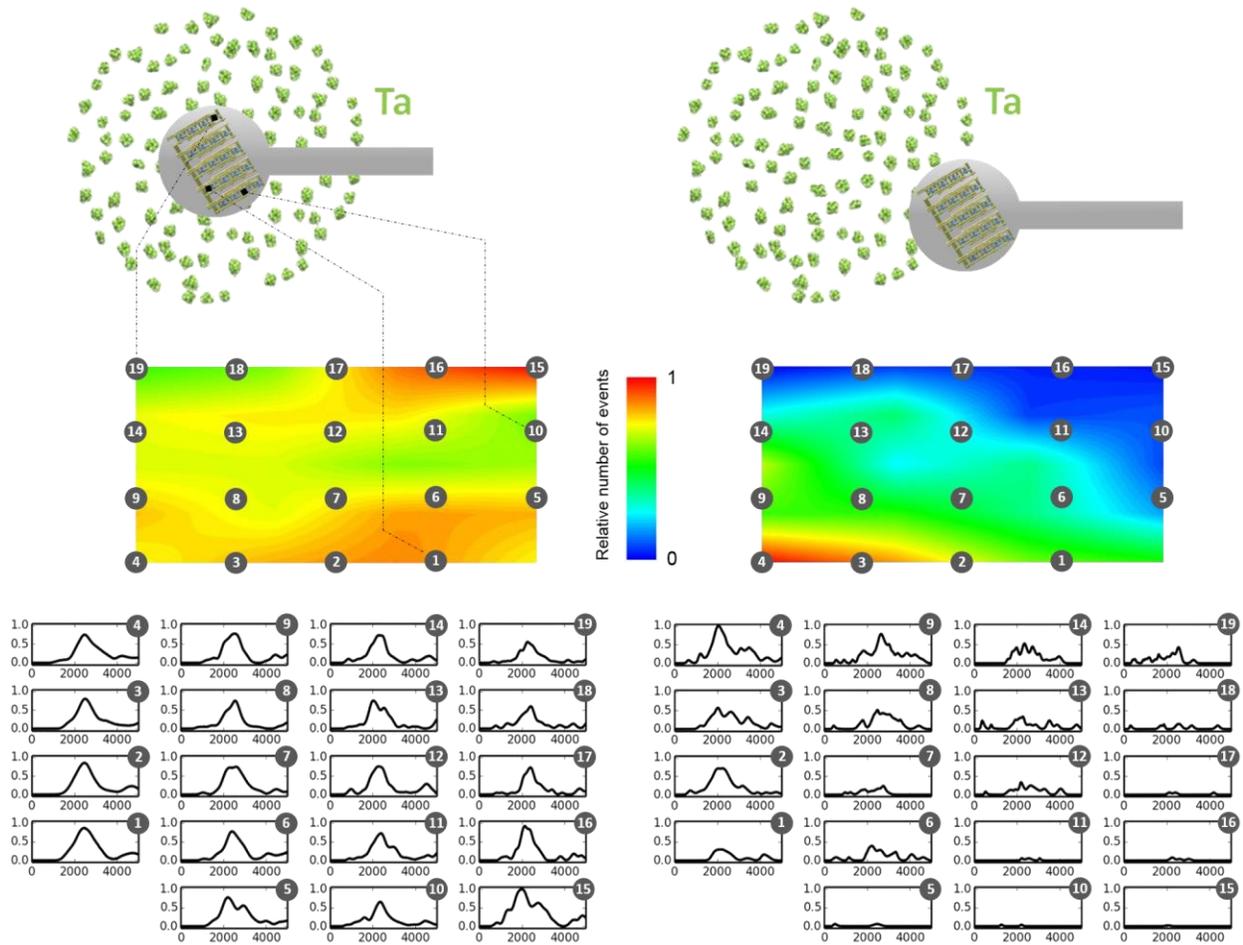

*Figure 4:* **NEMS-MS beam imaging**. *The NEMS array was placed at the center of the particle beam (left) or at the edge (right). The event number was measured for each NEMS and plotted on interpolated surface maps for each case. Mass spectra obtained with each individual resonator for a 4 minute acquisition are shown (bottom). A mechanical chopper was used to adapt the particle adsorption event rate to the array response time. The spectra are displayed as a matrix pattern reproducing the device physical layout (5×4). Each individual plot shows the intensity (a. u.) versus mass (kDa).*

In conclusion, we demonstrated single-particle nanomechanical mass spectrometry with arrays of NEMS operated with a frequency addressing scheme. These arrays can comprise several tens of nanoresonators, increasing the total capture cross-section by the same factor. Detection efficiency is today the main limitation of NEMS-based mass spectrometry[14] with analysis time up to several hours[13]. This time can be reduced by more than an order of magnitude with the frequency addressing technique, while keeping the same mass limit of detection: frequency stability of the silicon nanoresonators used here being limited by frequency fluctuations, it was not degraded by the frequency addressing technique. The number of resonators in the array could be further increased, with the caveat that both total input impedance of the array and output voltage would decrease accordingly. Additive white noise would eventually dominate and the frequency stability would degrade. We estimate that this may not be the case for arrays including up to between 50 to 100 nanoresonators. The main price to pay for the frequency addressing technique is an increase in duty cycle: an array comprising 100 resonators could be sampled in a few 100 ms. This is not very relevant for today's low-efficiency NEMS-MS systems, where the probability of multiple events within a duty cycle is very low. In the future however, as system particle transfer efficiencies improve, this probability will certainly increase. To circumvent such problem, several arrays with frequency addressing could eventually be operated simultaneously in parallel with multiple-channel electronics.

We have also demonstrated here how the frequency addressing scheme can provide individual resonator information: the array becomes a sort of particle imager, each resonator acting as a pixel. With this technique, Mass Spectrometry Imaging (MSI) at the single particle level becomes possible. MSI generally relies on the analysis of localized desorption events sequentially in time and has already proven its great potential for clinical applications and cancer research[19]. More recent techniques perform multi-pixel images with one single shot and efforts in the field are pushing towards better resolution imagers[20] as well as high-mass capability[21]. NEMS-MS imaging with frequency-addressed arrays has this potential. Ultimately, arrays of nanoresonators with µm-sized pixels covering large areas could be fabricated with CMOS co-integration[22]. Many frequency-addressed arrays could thus be operated simultaneously with an integrated electronics, like a CMOS imager. Beside biological research and biomedical applications, NEMS-MS imaging could be of great interest for the characterization of ionization sources efficiency, as well as the characterization of sampler performance in aerosol science.

AUTHOR INFORMATION

**Corresponding Author**

sebastien.hentz@cea.fr

**Present Addresses**

† Present address: APIX Analytics, 7 parvis Louis Néel – CS20050, 38040 Grenoble cedex 09, France

ACKNOWLEDGMENT

The authors acknowledge support from the LETI Carnot Institute NEMS-MS project, as well as from the European Union through the ERC Enlightened project (616251) and the Marie-Curie Eurotalents incoming (M.S.) fellowship.